\documentclass[technote,9pt]{./IEEEtran}
\usepackage[dvipdfm]{graphicx}
\usepackage[dvipdfm]{color}
\usepackage{amssymb}
\usepackage{amsmath}
\usepackage{subfigure}

\newtheorem{theorem}{Theorem}
\newtheorem{lemma}{Lemma}
\newtheorem{proposition}{Proposition}
\newtheorem{corollary}{Corollary}
\newtheorem{example}{Example}

\newcommand{\eps}{\epsilon}

\newcommand{\ticone}{\frac{2}{3}}
\newcommand{\tictwo}{H\left(\frac{1}{3}\right)-\frac{1}{3}}
\newcommand{\half}{\frac{1}{2}}

\begin{document}

\title{State Amplification}

\author{Young-Han Kim, Arak Sutivong, and Thomas M. Cover%
\thanks{Email: yhk@ucsd.edu, arak\_sutivong@mckinsey.com, cover@stanford.edu}}

\date{}

\maketitle

\begin{abstract}
We consider the problem of transmitting data at rate $R$ over a state
dependent channel $p(y|x,s)$ with state information available at the
sender and at the same time conveying the information about the
channel state itself to the receiver.  The amount of state information
that can be learned at the receiver is captured by the mutual
information $I(S^n; Y^n)$ between the state sequence $S^n$ and the
channel output $Y^n$.  The optimal tradeoff is characterized between
the information transmission rate $R$ and the state uncertainty
reduction rate $\Delta$, when the state information is either causally
or noncausally available at the sender. In particular, when state
transmission is the only goal, the maximum uncertainty reduction rate
is given by $\Delta^* = \max_{p(x|s)} I(X, S; Y)$.  This result is
closely related and in a sense dual to a recent study by Merhav and
Shamai, which solves the problem of \emph{masking} the state
information from the receiver rather than {conveying} it.
\end{abstract}

\section{Introduction}
\label{sec:intro}
A channel $p(y|x,s)$ with noncausal state information at the sender
has capacity
\begin{equation}
\label{eq:capacity}
C = \max_{p(u,x|s)} (I(U;Y) - I(U;S))
\end{equation}
as shown by Gelfand and Pinsker~\cite{Gelfand--Pinsker1980a}.
Transmitting at capacity, however, obscures the state information
$S^n$ as received by the receiver $Y^n$.  In some instances we wish to
convey the state information $S^n$ itself, which could be time-varying
fading parameters or an original image that we wish to enhance.  For
example, a stage actor with face $S$ uses makeup $X$ to communicate to
the back row audience $Y$. Here $X$ is used to enhance and exaggerate
$S$ rather than to communicate new information. Another motivation
comes from cognitive radio systems~\cite{FCC-Cognitive-Radio,
Mitola2000, Devroye--Mitran--Tarokh2006, Jovicic--Viswanath2006} with
the additional assumption that the secondary user $X^n$ communicates
its own message and at the same time facilitates the transmission of
the primary user's signal $S^n$. How should the transmitter
communicate over the channel to ``amplify'' his knowledge of the state
information to the receiver?  What is the optimal tradeoff between
state amplification and independent information transmission?

\begin{figure}[t]
\begin{center}
\input{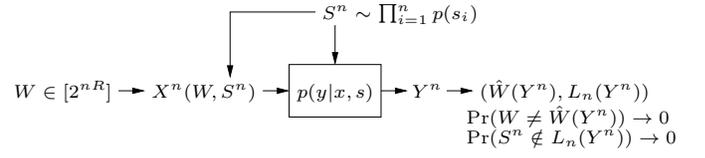}
\caption{Pure information transmission versus state uncertainty
reduction.}
\label{fig:setup}
\end{center}
\end{figure}

To answer these questions, we study the communication problem depicted
in Figure~\ref{fig:setup}.
Here the sender has access to the channel state sequence $S^n = (S_1,
S_2, \ldots, S_n)$, independent and identically distributed (i.i.d.)
according to $p(s)$, and wishes to transmit a message index $W \in
[2^{nR}] := \{1, 2, \ldots, 2^{nR}\}$, independent of $S^n$, as well
as to help the receiver reduce the uncertainty about the channel state
in $n$ uses of a state dependent channel $(\mathcal{X} \times
\mathcal{S}, p(y|x,s), \mathcal{Y})$.  Based on the message $W$ and
the channel state $S^n$, the sender chooses $X^n(W,S^n)$ and transmits
it across the channel. Upon observing the channel output $Y^n$, the
receiver guesses $\hat{W} \in [2^{nR}]$ and forms a list $L_n(Y^n)
\subseteq {\cal{S}}^n$ that contains likely candidates of the actual
state sequence $S^n$.

Without any observation $Y^n$, the receiver would know only that the
channel state $S^n$ is one of $2^{nH(S)}$ typical sequences (with
almost certainty) and we can say the uncertainty about $S^n$ is
$H(S^n)$. Now upon observing $Y^n$ and forming a list $L_n(Y^n)$ of
likely candidates for $S^n$, the receiver's list size is reduced from
$nH(S)$ to $\log |L_n|$. Thus we define the \emph{channel state
uncertainty reduction rate} to be
\[
\Delta = \frac{1}{n} \left (H(S^n)-\log|L_n|\right) 
= H(S) - \frac{1}{n}\log |L_n|
\]
as a natural measure for the amount of information the receiver learns
about the channel state.  In other words, the uncertainty reduction
rate $\Delta \in [0, H(S)]$ captures the difference between the
original channel state uncertainty and the residual state uncertainty
after observing the channel output. Later in
Section~\ref{sec:gaussian} we will draw a connection between the list
size reduction and the conventional information measure $I(S^n; Y^n)$
that also captures the amount of information $Y^n$ learns about $S^n$.

More formally, we define a $(2^{nR}, 2^{n\Delta}, n)$ code as the
encoder map
\[
X^n: [2^{nR}] \times \mathcal{S}^n \to \mathcal{X}^n
\]
and decoder maps
\begin{align*}
\hat{W} &: \mathcal{Y}^n \to [2^{nR}]\\
L_n &: \mathcal{Y}^n \to 2^{\mathcal{S}^n}
\end{align*}
with list size 
\[
|L_n| = 2^{n(H(S) - \Delta)}.
\]
The probability of a
message decoding error $P_{e,w}^{(n)}$ and the probability of a list
decoding error $P_{e,s}^{(n)}$ are defined respectively as
\begin{align*}
P^{(n)}_{e,w} &= \frac{1}{2^{nR}}\sum_{w=1}^{2^{nR}}
\Pr(\hat{W} \neq w|W=w),\\
P^{(n)}_{e,s} &= \Pr(S^n \notin L_n(Y^n))
\end{align*}
where the message index $W$ is chosen uniformly over $[2^{nR}]$ and
the state sequence $S^n$ is drawn i.i.d.\@ $\sim p(s)$, independent of
$W$.  A pair $(R,\Delta)$ is said to be achievable if there exists a
sequence of $(2^{nR}, 2^{n\Delta}, n)$ codes with $P^{(n)}_{e,w}
\rightarrow 0$ and $P^{(n)}_{e,s} \rightarrow 0$ as $n \rightarrow
\infty$. Finally, we define the optimal $(R, \Delta)$ tradeoff region,
or the \emph{tradeoff region} in short, to be the closure of all
achievable $(R, \Delta)$ pairs, and denote it by $\mathcal{R}^*$.

This paper shows that the tradeoff region $\mathcal{R}^*$ can be
characterized as the union of all $(R,\Delta)$ pairs satisfying
\begin{align*}
R &\le I(U; Y) - I(U;S) \\
\Delta &\le H(S)\\
R + \Delta &\le I(X, S; Y)
\end{align*}
for some joint distribution of the form $p(s)p(u,x|s)p(y|x,s)$.

As a special case, if the encoder's sole goal is to ``amplify'' the
state information ($R = 0$), then the maximum uncertainty reduction
rate
\[
\Delta^* = \sup
\{\Delta: (R, \Delta) \text{ is achievable for some $R \ge 0$}\}
\]
is given by
\begin{equation}
\label{eq:max-delta}
\Delta^* = \min \{ H(S), \enspace \max_{p(x|s)} I(X, S; Y) \}.
\end{equation}
The maximum uncertainty reduction rate $\Delta^*$ is achieved by
designing the signal $X^n$ to enhance the receiver's estimation of the
state $S^n$ while using the remaining pure information bearing freedom
in $X^n$ to provide more information about the state.  More
specifically, there are three different components involved in
reducing the receiver's uncertainty about the state:
\begin{enumerate}
\item[1)] The transmitter uses the channel capacity to convey the
state information.  In Section~\ref{sec:noncausal}, we study the
classical setup~\cite{Kuznetsov--Tsybakov1974, Heegard--El-Gamal1983}
of coding for memory with defective cells (Example~\ref{ex:memory})
and show that this ``source-channel separation'' scheme is optimal
when the memory defects are symmetric.

\item[2)] The transmitter gets out of the way of the receiver's view
of the state.  For instance, the maximum uncertainty reduction for the
binary multiplying channel $Y=X \cdot S$ (Example~\ref{ex:binary} in
Section~\ref{sec:noncausal}) with binary input $X \in \{0,1\}$ and
binary state $S \in \{0, 1\}$ is achieved by sending $X \equiv 1$.

\item[3)] The transmitter actively amplifies the state.  In
Example~\ref{ex:wdp} in Section~\ref{sec:gaussian}, we consider the
Gaussian channel $Y = X + S + Z$ with Gaussian state $S$ and Gaussian
noise $Z$.  Here the optimal transmitter amplifies the state as $X =
\alpha S$ under the given power constraint $E X^2 \le P$.

\end{enumerate}

It is interesting to note that the maximum uncertainty reduction rate
$\Delta^*$ is the information rate $I(X,S; Y)$ that could be achieved
if both the state $S$ and the signal $X$ could be freely designed,
instead of the state $S$ being generated by nature.  This rate also
appears in the sum rate of the capacity region expression for the
cooperative multiple access channel~\cite[Problem
15.1]{Cover--Thomas2006} and the multiple access channel with cribbing
encoders by Willems and van der
Meulen~\cite{Willems--van-der-Meulen1985}.

When the state information is only \emph{causally} available at the
transmitter, that is, when the channel input $X_i$ depends on only the
past and the current channel channel state $S^i$, we will show that
the tradeoff region $\mathcal{R}^*$ is given as the union of all
$(R,\Delta)$ pairs satisfying
\begin{align*}
R &\le I(U; Y) \\
\Delta &\le H(S)\\
R + \Delta &\le I(X, S; Y)
\end{align*}
over all joint distributions of the form $p(s)p(u)p(x|u,s)p(y|x,s)$.
Interestingly, the maximum uncertainty reduction rate $\Delta^*$ stays
the same as in the noncausal case~\eqref{eq:max-delta}.  That
causality incurs no cost on the (sum) rate is again reminiscent of the
multiple access channel with cribbing
encoders~\cite{Willems--van-der-Meulen1985}.

The problem of communication over state-dependent channels with state
information known at the sender has attracted a great deal of
attention.  This research area was first pioneered by
Shannon~\cite{Shannon1958a}, Kuznetsov and
Tsybakov~\cite{Kuznetsov--Tsybakov1974}, and Gelfand and
Pinsker~\cite{Gelfand--Pinsker1980a}.  Several advancements in both
theory and practice have been made over the years.  For instance,
Heegard and El Gamal~\cite{Heegard--El-Gamal1983, Heegard1981}
characterized the channel capacity and devised practical coding
techniques for computer memory with defective cells.  Costa
\cite{Costa1983} studied the now famous ``writing on dirty paper''
problem and showed that the capacity of an additive white Gaussian
noise channel is not affected by additional interference, as long as
the entire interference sequence is available at the sender prior to
the transmission.  This fascinating result has been further extended
with strong motivations from applications in digital watermarking
(see, for example, Moulin and O'Sullivan~\cite{Moulin--OSullivan2003},
Chen and Wornell~\cite{Chen--Wornell2001}, and Cohen and
Lapidoth~\cite{Cohen--Lapidoth2002}) and multi-antenna broadcast
channels (see, for example, Caire and Shamai~\cite{Caire--Shamai2003},
Weingarten, Steinberg, and
Shamai~\cite{Weingarten--Steinberg--Shamai2006}, and Mohseni and
Cioffi~\cite{Mohseni--Cioffi2006}).  Readers are referred to Caire and
Shamai~\cite{Caire--Shamai1999}, Lapidoth and
Narayan~\cite{Lapidoth--Narayan1998}, and Jafar~\cite{Jafar2006} for
more complete reviews on the theoretical development of the field.  On
the practical side, Erez, Shamai, and
Zamir~\cite{Erez--Shamai--Zamir2005, Zamir--Shamai--Erez2002} proposed
efficient coding schemes based on lattice strategies for binning.
More recently, Erez and ten Brink~\cite{Erez--ten-Brink2005} report
efficient coding techniques that almost achieve the capacity of
Costa's dirty paper channel.

In \cite{Sutivong--Chiang--Cover--Kim2005,
Sutivong--Cover--Chiang--Kim2002}, we formulated the problem of
simultaneously transmitting pure information and helping the receiver
estimate the channel state under a distortion measure.  Although the
characterization of the optimal rate-distortion tradeoff is still open
in general (cf.~\cite{Sutivong2003}), a complete solution is given for
the Gaussian case (the writing on dirty paper channel) under quadratic
distortion~\cite{Sutivong--Chiang--Cover--Kim2005}.  In this
particular case, optimality was shown for a simple power-sharing
scheme between pure information transmission via Costa's original
coding scheme and state amplification via simple scaling.

Recently, Merhav and Shamai~\cite{Merhav--Shamai2007} considered a
related problem of transmitting pure information, but this time under
the additional requirement of \emph{minimizing} the amount of
information the receiver can learn about the channel state.  In this
interesting work, the optimal tradeoff between pure information rate
$R$ and the amount of state information $E$ is characterized for both
causal and noncausal setups.  Furthermore, for the Gaussian noncausal
case (writing on dirty paper), the optimal rate-distortion tradeoff is
given under quadratic distortion.  (This may well be called ``writing
dirty on paper''.)

The current paper thus complements \cite{Merhav--Shamai2007} in a dual
manner.  It is refreshing to note that our notion of uncertainty
reduction rate $\Delta$ is essentially equivalent to Merhav and
Shamai's notion of $E$; both notions capture the normalized mutual
information $I(S^n;Y^n)$.  (See the discussion in
Section~\ref{sec:gaussian}.)  The crucial difference is that $\Delta$
is to be maximized while $E$ is to be minimized.  Both problems admit
single-letter optimal solutions. 


The rest of this paper is organized as follows.  In the next section,
we establish the optimal $(R, \Delta)$ tradeoff region for the case in
which the state information $S^n$ is noncausally available at the
transmitter before the actual communication.
Section~\ref{sec:gaussian} extends the notion of state uncertainty
reduction to continuous alphabets, by identifying the list decoding
requirement ${S}^n \in L_n(Y^n)$ with the mutual information rate
$\frac{1}{n} I(S^n; Y^n)$.  In particular, we characterize the optimal
$(R,\Delta)$ tradeoff region for Costa's ``writing on dirty paper''
channel.  Since the intuition gained from the study of the noncausal
setup carries over when the transmitter has causal knowledge of the
state sequence, the causal case is treated only briefly in
Section~\ref{sec:causal}, followed by concluding remarks in
Section~\ref{sec:conc}.

\section{Optimal $(R,\Delta)$ Tradeoff: Noncausal Case}
\label{sec:noncausal}

In this section, we characterize the optimal tradeoff region between
the pure information rate $R$ and the state uncertainty reduction rate
$\Delta$ with state information noncausally available at the
transmitter, as formulated in Section~\ref{sec:intro}.

\begin{theorem}
\label{thm:noncausal}
The tradeoff region $\mathcal{R}^*$ for a state-dependent channel
$(\mathcal{X} \times \mathcal{S}, p(y|x,s), \mathcal{Y})$ with 
state information $S^n$ noncausally known at the transmitter is the
union of all $(R, \Delta)$ pairs satisfying
\begin{align}
R &\le I(U; Y) - I(U;S) \label{eq:noncausal1}\\
\Delta &\le H(S) \label{eq:noncausal2}\\
R + \Delta &\le I(X, S; Y) \label{eq:noncausal3}
\end{align}
for some joint distribution of the form $p(s)p(u,x|s)p(y|x,s)$,
where the auxiliary random variable $U$ has cardinality bounded by
$|\mathcal{U}| \le |\mathcal{X}| \cdot |\mathcal{S}|$.
\end{theorem}

As will be clear from the proof of the converse, the region given by
\eqref{eq:noncausal1}--\eqref{eq:noncausal3} is convex.  (We can merge
the time-sharing random variable into $U$.)  Since the auxiliary
random variable $U$ affects the first inequality \eqref{eq:noncausal1}
only, the cardinality bound on $\mathcal{U}$ follows directly from the
usual technique; see Gelfand and Pinsker~\cite{Gelfand--Pinsker1980a}
or a general treatment by Salehi~\cite{Salehi1978}.  Finally, we can
take $X$ as a deterministic function of $(U,S)$ without reducing the
region, but at the cost of increasing the cardinality bound of $U$;
refer to the proof of Lemma~\ref{lemma:equiv} below.

It is easy to see that we can recover the Gelfand--Pinsker capacity
formula
\begin{align*}
C &= \max \{R: (R,\Delta) \in \mathcal{R}^* \text{ for some $\Delta \ge 0$}\}\\
&= \max_{p(x,u|s)} ( I(U;Y) - I(U;S) ).
\end{align*}
For the other extreme case of pure state amplification, we have the
following result.
\medskip
\begin{corollary}
\label{corollary:noncausal}
Under the condition of Theorem~\ref{thm:noncausal}, the maximum
uncertainty reduction rate $\Delta^*
= \max \{\Delta: (R, \Delta) \in \mathcal{R}^* 
\text{ for} \text{ some}\linebreak R \ge 0\}$
is given by
\begin{equation}
\label{eq:max-delta2}
\Delta^* 
= \min \{ H(S), \enspace \max_{p(x|s)} I(X, S; Y) \}.
\end{equation}
\end{corollary}

Thus the receiver can learn about the state $S^n$ essentially at the
maximal cut-set rate $I(X, S; Y).$

Before we prove Theorem~\ref{thm:noncausal}, we need the following two
lemmas.  The first one extends Fano's inequality~\cite[Lemma
7.9.1]{Cover--Thomas2006} to list decoding.
\medskip
\begin{lemma}
\label{lemma:gen-fano}
For a sequence of list decoders $L_n: \mathcal{Y}^n \to 2^{\mathcal{S}^n},$
$Y^n \mapsto L_n(Y^n)$ with list size $|L_n|$ fixed for each 
$n$, let $P^{(n)}_{e,s} = \Pr(S^n \notin L_n(Y^n))$ be
the sequence of corresponding probabilities of list decoding error.
If $P_{e,s}^{(n)} \to 0$, then
\[
H(S^n|Y^n) \le \log |L_n| + n\eps_n
\]
where $\eps_n \to 0$ as $n \to \infty$.
\end{lemma}
\begin{IEEEproof}
Define an error random variable $E$ as
\[
E = \left\{
\begin{array}{ll} 
0, & \text{if $S^n \in L_n,$} \\ 
1, & \text{if $S^n \notin L_n.$}
\end{array}\right.
\]
We can then expand 
\begin{align*}
H(E,S^n|Y^n) &= H(S^n|Y^n) + H(E|Y^n,S^n)\\
&= H(E|Y^n) + H(S^n|Y^n,E).
\end{align*}
Note that $H(E|Y^n) \le 1$ and $H(E|Y^n,S^n) = 0$. We can also bound
$H(S^n|Y^n,E)$ as
\begin{align*}
H(S^n|E,Y^n) &= H(S^n|Y^n,E=0)\mbox{Pr}(E=0) \\
&\quad + H(S^n|Y^n,E=1)\mbox{Pr}(E=1)\\ 
&\le      \log|L_n|(1-P^{(n)}_{e,s}) +
          n\log|{\cal{S}}|P^{(n)}_{e,s}
\end{align*}
where the inequality follows because when there is no error, the
remaining uncertainty is at most $\log|L_n|$, and when there is an
error, the uncertainty is at most $n\log|{\cal{S}}|$.  This implies
that
\begin{align*}
H(S^n|Y^n) &\le 1+\log|L_n|(1-P^{(n)}_{e,s}) + n\log|{\cal{S}}|P^{(n)}_{e,s}\\
           &=   \log|L_n| + 1 + (n\log|{\cal{S}}| - \log|L_n|)P^{(n)}_{e,s}.
\end{align*}
Taking $\epsilon_n = \frac{1}{n} + (\log|{\cal{S}}| -
\frac{1}{n}\log|L_n|)P^{(n)}_{e,s}$ proves the desired result.
\end{IEEEproof}

The second lemma is crucial to the proof of
Theorem~\ref{thm:noncausal} and contains a more interesting technique
than Lemma~\ref{lemma:gen-fano}. This lemma shows that the third
inequality \eqref{eq:noncausal3} can be replaced by a tighter
inequality \eqref{eq:noncausal4} below (recall that $I(U,S;Y) \le
I(X,S;Y)$ since $U \to (X,S) \to Y$), which becomes crucial
for the achievability
proof of Theorem~\ref{thm:noncausal}.
\medskip
\begin{lemma}
\label{lemma:equiv}
Let $\mathcal{R}$ be the union of all $(R,\Delta)$ pairs satisfying
\eqref{eq:noncausal1}--\eqref{eq:noncausal3}.  Let ${\mathcal{R}}_0$
be the {closure} of the union of all $(R,\Delta)$ pairs satisfying
\begin{align}
R &\le I(U; Y) - I(U;S) \tag{\ref{eq:noncausal1}}\\
\Delta &\le H(S) \tag{\ref{eq:noncausal2}}\\
\label{eq:noncausal4}
R + \Delta &\le I(U, S; Y)
\end{align}
for some joint distribution $p(s) p(x,u|s) p(y|x,s)$, where the
auxiliary random variable $U$ has finite cardinality.  Then
\[
\mathcal{R} = {\mathcal{R}}_0.
\]
\end{lemma}
\begin{IEEEproof}
Since $U \to (X,S) \to Y$ forms a Markov chain, it is trivial to check
that 
\begin{equation}
\label{eq:inclusion1}
\mathcal{R}_0 \subseteq {\mathcal{R}}.
\end{equation}
For the other direction of inclusion, we need some notation.
Let $\mathcal{P}$ be the set of all distributions of the form $p(s)
p(x,u|s) p(y|x,s)$ consistent with the given $p(s)$ and $p(y|x,s)$,
where the auxiliary random variable $U$ is defined on an arbitrary
finite set.  Further let $\mathcal{P}'$ be the restriction of
$\mathcal{P}$ such that $X = f(U,S)$ for some function $f$, i.e.,
$p(x|u,s)$ takes values $0$ or $1$ only.

If we define $\mathcal{R}_1$ to denote the closure of all $(R,\Delta)$
pairs satisfying \eqref{eq:noncausal1}, \eqref{eq:noncausal2}, and
\eqref{eq:noncausal4} over $\mathcal{P}'$, or equivalently, if
$\mathcal{R}_1$ is defined to be the restriction of $\mathcal{R}_0$
over a smaller set of distributions $\mathcal{P}'$, then clearly
\begin{equation}
\label{eq:inclusion2}
\mathcal{R}_1 \subseteq \mathcal{R}_0.
\end{equation}
Let $\mathcal{R}_2$ be defined as the closure of $(R, \Delta)$
pairs satisfying \eqref{eq:noncausal1}--\eqref{eq:noncausal3}.  Since
$X \to (U,S) \to Y$ forms a Markov chain on $\mathcal{P}'$, we
have
\begin{equation}
\label{eq:inclusion3}
\mathcal{R}_2 \subseteq \mathcal{R}_1.
\end{equation}

To complete the proof, it now suffices to show that 
\begin{equation}
\label{eq:inclusion4}
\mathcal{R} \subseteq \mathcal{R}_2.
\end{equation}
To see this, we restrict $\mathcal{R}_2$ to the distributions of the
form $U = (V, \tilde{U})$ with $V$ independent of $(\tilde{U}, S)$, namely,
\begin{equation}
\label{eq:new-dist}
p(x,u|s) = p(x, v, \tilde{u} | s) = p(v) p(\tilde{u} | s) p(x|v,
\tilde{u}, s)
\end{equation}
with deterministic $p(x|v,\tilde{u},s)$, i.e., $x$ is a function of
$(v,\tilde{u},s)$, and call this restriction $\mathcal{R}_3$.  Since
$X$ is a deterministic function of $(V, \tilde{U}, S)$ and at the same
time $(V, \tilde{U}) \to (X,S) \to Y$ form a Markov chain,
$\mathcal{R}_3$ can be written as the closure of all $(R, \Delta)$
pairs satisfying
\begin{align*}
R &\le I(V, \tilde{U}; Y) - I(V, \tilde{U}; S) \\
\Delta &\le H(S) \\
R + \Delta &\le I(V, \tilde{U}, S; Y) = I(X, S; Y)
\end{align*}
for some distribution of the form $p(s)p(x,v,\tilde{u}|s)p(y|x,s)$
satisfying \eqref{eq:new-dist}.  But we have
\begin{align*}
\nonumber
I(V, \tilde{U}; Y) - I(V, \tilde{U}; S) 
&\ge I(\tilde{U}; Y) - I(V, \tilde{U}; S)\\
&= I(\tilde{U}; Y) - I(\tilde{U}; S)
\end{align*}
and the set of conditional distributions on $(\tilde{U}, X)$ given $S$
satisfying \eqref{eq:new-dist} is as rich as any $p(\tilde{u}, x |
s)$.  (Indeed, any conditional distribution $p(a|b)$ can be
represented as $\sum_c p(c)p(a|b,c)$ for appropriately chosen $p(c)$
and \emph{deterministic} distribution $p(a|b,c)$ with cardinality of
$C$ upper bounded by $(|\mathcal{A}|-1)|\mathcal{B}| + 1$; see also
\cite[Eq.~(44)]{Willems--van-der-Meulen1985}.)  Therefore, we have
\begin{equation}
\label{eq:inclusion5}
\mathcal{R} \subseteq \mathcal{R}_3 \subseteq \mathcal{R}_2
\end{equation}
which completes the proof.
\end{IEEEproof}

Now we are ready to prove Theorem~\ref{thm:noncausal}.
\begin{IEEEproof}[Proof of Theorem~\ref{thm:noncausal}]
For the proof of achievability, in the light of
Lemma~\ref{lemma:equiv}, it suffices to prove that any pair $(R,
\Delta)$ satisfying \eqref{eq:noncausal1}, \eqref{eq:noncausal2},
\eqref{eq:noncausal4} for some $p(u,x|s)$ is achievable.  Since the
coding technique is quite standard, we only sketch the proof here.
For fixed $p(u,x|s)$, the result of
Gelfand--Pinsker~\cite{Gelfand--Pinsker1980a} shows that the
transmitter can send $I(U;Y) - I(U;S)$ bits reliably across the
channel.  Now we allocate $0 \le R \le I(U;Y) - I(U;S)$ bits for
sending the pure information and use the remaining $\Gamma = I(U;Y) -
I(U;S) - R$ bits for sending the state information by random
binning. More specifically, we assign typical $S^n$ sequences to
$2^{n\Gamma}$ bins at random and send the bin index of the observed
$S^n$ using $n \Gamma$ bits.  At the receiving end, the receiver is
able to decode the codeword $U^n$ from $Y^n$ with high probability.
Using joint typicality of $(Y^n, U^n, S^n)$, the state uncertainty can
be first reduced from $H(S)$ to $H(S|Y,U)$. Indeed, the number of
typical $S^n$ sequences jointly typical with $(Y^n, U^n)$ is bounded
by $2^{n (H(S|Y,U) + \eps)}$. In addition, using $\Gamma = I(U;Y) -
I(U;S) - R$ bits of independent refinement information from the hash
index of $S^n$, we can further reduce the state uncertainty by
$\Gamma$. Hence, by taking the list of all $S^n$ sequences jointly
typical with $(Y^n, U^n)$ satisfying the hash check, we have the total
state uncertainty reduction rate 
\begin{align*}
\Delta &= I(U,Y;S) + \Gamma \\
&= I(U,Y;S) + I(U;Y) - I(U;S) - R \\
&= I(U,S;Y) - R.
\end{align*}
By varying $0 \le R \le I(U;Y) - I(U;S)$, it can be readily seen that
all $(R,\Delta)$ pairs satisfying
\begin{align*}
R &\le I(U;Y) - I(U;S)\\
\Delta &\le H(S)\\
R + \Delta &\le I(U,S;Y)
\end{align*}
for any fixed $p(x,u|s)$ are achievable.

For the proof of converse, we have to show that given any sequence of
$(2^{nR}, 2^{n\Delta}, n)$ codes with $P^{(n)}_{e,w}, P^{(n)}_{e,s}
\to 0,$ the $(R, \Delta)$ pairs must satisfy 
\begin{align*}
R &\le I(U;Y) - I(U;S) \\
\Delta &\le H(S) \\
R + \Delta &\le I(X,S;Y)
\end{align*}
for some joint distribution $p(s)p(x,u|s)p(y|x,s)$.

The pure information rate $R$ can be readily bounded from the previous
work by Gelfand and Pinsker~\cite[Proposition
3]{Gelfand--Pinsker1980a}.  Here we repeat a simpler proof given in
Heegard~\cite[Appendix 2]{Heegard1981} for completeness; see also
\cite[Lecture~13]{El-Gamal2006}.  Starting with Fano's inequality, we
have the following chain of inequalities:
\begin{align*}
n R &\leq  I(W;Y^n)+n\epsilon_n \\
&= \sum_{i=1}^n I(W;Y_i|Y^{i-1})+n\epsilon_n \\
&\leq  \sum_{i=1}^n I(W,Y^{i-1};Y_i)+n\epsilon_n \\
&= \sum_{i=1}^n I(W,Y^{i-1}\!,S_{i+1}^n;Y_i) 
   -\sum_{i=1}^n I(Y_i;S_{i+1}^n|W,Y^{i-1}) +n\epsilon_n \\
&\stackrel{\text{(a)}}{=} 
\sum_{i=1}^n I(W,Y^{i-1}\!,S_{i+1}^n;Y_i) 
   -\sum_{i=1}^n I(Y^{i-1};S_i|W,S_{i+1}^n) +n\epsilon_n \\
&\stackrel{\text{(b)}}{=} \sum_{i=1}^n I(W,Y^{i-1}\!,S_{i+1}^n;Y_i) 
   -\sum_{i=1}^n I(W,Y^{i-1}\!,S_{i+1}^n;S_i) +n\epsilon_n
\end{align*}
where (a) follows from the Csisz\'ar sum formula
\begin{align*}
\sum_{i=1}^n I(Y_i;S_{i+1}^n|W,Y^{i-1}) &= 
\sum_{i=1}^n \sum_{j=i+1}^n I(Y_i; S_j | W, S_{j+1}^n, Y^{i-1})\\
&= \sum_{j=1}^n \sum_{\hspace{5.25pt}i=1\hspace{5.25pt}}^{j-1} 
    I(Y_i; S_j | W, S_{j+1}^n, Y^{i-1})\\
&= \sum_{j=1}^n I(Y^{j-1}; S_j | W, S_{j+1}^n)
\end{align*}
and (b) follows because $(W,S_{i+1}^n)$ is independent of $S_i$.  By
recognizing the auxiliary random variable $U_i = (W, Y^{i-1},
S_{i+1}^n)$ and noting that $U_i \to (X_i, S_i) \to Y_i$ form a Markov
chain, we have
\begin{equation}
\label{eq:before-q1}
nR \le \sum_{i=1}^n (I(U_i;Y_i) - I(U_i;S_i)) + n\eps_n.
\end{equation}

On the other hand, since $\log |L_n| = n (H(S) - \Delta)$, we can
trivially bound $\Delta$ by Lemma~\ref{lemma:gen-fano} as
\begin{align*}
n\Delta &\le n H(S) - H(S^n|Y^n) + n\epsilon'_n \nonumber\\
       &\le n H(S) + n \epsilon'_n.
\end{align*}
Similarly, we can bound $R+\Delta$ as
\begin{align}
n(R+\Delta)
&\le I(W;Y^n)+I(S^n;Y^n) + n\epsilon''_n\nonumber\\
&\stackrel{\text{(a)}}{\le} I(W;Y^n|S^n)+I(S^n;Y^n) +
n\epsilon''_n\nonumber\\
&\le I(W,S^n;Y^n) + n\epsilon''_n \nonumber\\
&\stackrel{\text{(b)}}{=} I(X^n,S^n;Y^n) + n \epsilon''_n \nonumber\\
&\stackrel{\text{(c)}}{\le} \frac{1}{n} \sum_{i=1}^n I(X_i,S_i; Y_i) +
\epsilon''_n \label{eq:before-q2}
\end{align}
where (a) follows since $W$ is independent of $S^n$ and conditioning
reduces entropy, (b) follows from the data processing inequality
(both directions), and (c) follows from the memorylessness of the
channel.

We now introduce the usual time-sharing random variable $Q$ uniform
over $\{1,\ldots, n\}$, independent of everything else.  Then
\eqref{eq:before-q1} implies
\begin{align*}
R &\le I(U_Q; Y_Q | Q) - I(U_Q; S_Q | Q) + \eps_n \\
&= I(U_Q, Q; Y_Q) - I(U_Q, Q; S_Q) + \eps_n.
\end{align*}
On the other hand, \eqref{eq:before-q2} implies
\begin{align*}
R + \Delta &\le I(X_Q, S_Q; Y_Q | Q) + \eps''_n \\
&\le I(X_Q, S_Q, Q; Y_Q) + \eps''_n \\
&= I(X_Q, S_Q; Y_Q) + \eps''_n
\end{align*}
where the last equality follows since $Q \to (X_Q, S_Q) \to Y_Q$ form
a Markov chain.

Finally, we recognize $U = (U_Q, Q), X = X_Q, S = S_Q,$ $Y = Y_Q,$ and
note that $S \sim p(s)$, $\Pr(Y =y | X = x, S = s) = p(y|x,s)$, and $U
\to (X,S) \to Y$, which completes the proof of the converse.
\end{IEEEproof}

\medskip

Roughly speaking, the optimal coding scheme is equivalent to sending
the codeword $U^n$ reliably at the Gelfand--Pinsker rate $R' = I(U;Y)
- I(U;S)$ and reducing the receiver's uncertainty by $\Delta' = I(S;
U,Y)$ from $Y^n$ and the decoded codeword $U^n$.  It should be noted
that $(R',\Delta')$ has the same form as the achievable region for the
dual tradeoff problem between pure information rate $R$ and (minimum)
normalized mutual information rate $E = \frac{1}{n}I(S^n;Y^n)$ studied
in \cite{Merhav--Shamai2007}. But we can reduce the uncertainty about
$S^n$ further by allocating part $\Gamma$ of the pure information rate
$R'$ to convey independent refinement information (hash index of
$S^n$). By varying $\Gamma \in [0, R']$ we can trace the entire
tradeoff region $(R' - \Gamma, \Delta' + \Gamma)$.

It turns out an alternative coding scheme based on Wyner--Ziv source
coding with side information~\cite{Wyner--Ziv1976}, instead of random
binning, also achieves the tradeoff region $\mathcal{R}^*$.  To see
this, fix any $p(u,x|s)$ and $p(v|s)$ satisfying
\[
\Gamma := I(V;S|U,Y) \le I(U;Y) - I(U;S)
\]
and consider the Wyner--Ziv encoding of $S^n$ with covering codeword
$V^n$ and side information $(U^n, Y^n)$ at the decoder. More
specifically, we can generate $2^{nI(V;S)}$ $V^n$ codewords and assign
them into $2^{n\Gamma}$ bins.  As before we use the Gelfand--Pinsker
coding to convey a message of rate $I(U;Y) - I(U;S)$ reliably over the
channel. Since the rate $\Gamma = I(V; S | U,Y)$ is sufficient to
reconstruct $V^n$ at the receiver with side information $Y^n$ and
$U^n$, we can allocate the rate $\Gamma$ for conveying $V^n$ and use
the remaining rate $R = I(U;Y) - I(U;S) - \Gamma$ for extra pure
information. Forming a list of $S^n$ jointly typical with $(Y^n, U^n,
V^n)$ results in the uncertainty reduction rate $\Delta$ given by
\begin{align*}
\Delta &= I(S; Y, U, V) \\
&= I(S; Y, U) + \Gamma \\
&= I(S; U, Y) + I(U;Y) - I(U;S) - R\\
&= I(U, S; Y) - R.
\end{align*}

Thus the tradeoff region ${\mathcal{R}}^*$ can be achieved via the
combination of two fundamental results in communication with side
information: channel coding with side information by Gelfand and
Pinsker~\cite{Gelfand--Pinsker1980a} and rate distortion with side
information by Wyner and Ziv~\cite{Wyner--Ziv1976}.  It is also
interesting to note that the information about $S^n$ can be
transmitted in a manner completely independent of geometry (random
binning) or completely dependent on geometry (random covering); refer
to \cite{Cover--Kim2007} for a similar phenomenon in a relay channel
problem.

\medskip

When $Y$ is a function of $(X,S)$, it is optimal to
identify $U = Y,$ and Theorem~\ref{thm:noncausal} simplifies to the
following corollary.

\begin{corollary}
\label{coro:deterministic}
The tradeoff region $\mathcal{R}^*$ for a deterministic
state-dependent channel $Y = f(X,S)$ with state information $S^n$
noncausally known at the transmitter is the union of all $(R, \Delta)$
pairs satisfying
\begin{align}
R &\le H(Y|S) \label{eq:deterministic1}\\
\Delta &\le H(S) \label{eq:deterministic2}\\
R + \Delta &\le H(Y) \label{eq:deterministic3}
\end{align}
for some joint distribution of the form $p(s)p(x|s)p(y|x,s)$.  In
particular, the maximum uncertainty reduction rate is given by
\begin{equation}
\Delta^* = \min \{ H(S),\enspace \max_{p(x|s)} H(Y)\}.
\end{equation}
\end{corollary}

The next two examples show different flavors of optimal state
uncertainty reduction.

\medskip
\begin{example}
\label{ex:memory}
Consider the problem of conveying information using a write-once
memory device with stuck-at defective cells
\cite{Kuznetsov--Tsybakov1974, Heegard--El-Gamal1983} as depicted in
Figure~\ref{fig:memory}.
\begin{figure}[t]
\begin{center}
\input{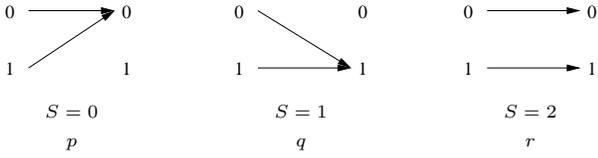}
\caption{Memory with defective cells.}
\label{fig:memory}
\end{center}
\end{figure}
Here each memory cell has probability $p$ of being stuck at $0$,
probability $q$ of being stuck at $1$, and probability $r$ of being a
good cell, with $p + q + r = 1$.  It is easy to see that the channel
output $Y$ is a simple deterministic function of the channel input $X$
and the state $S$.

Now it is easy to verify that the tradeoff region $\mathcal{R}^*$ is
given by
\begin{align}
R &\le r  H(\alpha) \label{eq:memory1}\\
\Delta &\le H(p,q,r) \label{eq:memory2}\\
R + \Delta &\le H(p + \alpha r, q + (1-\alpha) r) \label{eq:memory3}
\end{align}
where $\alpha$ can be chosen arbitrarily ($0 \le \alpha \le 1$).  This
region is achieved by choosing $p(x) \sim \text{Bern}(\alpha)$.
Without loss of generality, we can choose $X \sim \text{Bern}(\alpha)$
independent of $S$, because the input $X$ affects $Y$ only when $S=2$.

There are two cases to consider.
\begin{enumerate}
\item[(a)] If $p = q$, then the choice of $\alpha^* = 1/2$ maximizes
both \eqref{eq:memory1} and \eqref{eq:memory3}, and hence achieves the
entire tradeoff region $\mathcal{R}^*$.  The optimal transmitter
splits the full channel capacity $C = r H(\alpha^*) = r$ to send both
the pure information and the state information.  (See
Figure~\ref{fig:graph}(a) for the case $(p, q, r) = (1/3, 1/3, 1/3)$.)

\item[(b)] On the other hand, when $p \ne q$, there is a clear
tradeoff in our choice of $\alpha$.  For example, consider the case
$(p,q,r) = (1/2, 1/6, 1/3)$.  If the goal is to communicate pure
information over the channel, we should take $\alpha^* = 1/2$ to
maximize the number of distinguishable input preparations.  This gives
the channel capacity $C = r H(\alpha) = 1/3$.  If the goal is,
however, to help the receiver reduce the state uncertainty, we take
$\alpha^* = 0$, i.e., we transmit a fixed signal $X \equiv 0$.  This
way, the transmitter can minimize his interference with the receiver's
view of the state $S$.  The entire tradeoff region is given in
Figure~\ref{fig:graph}(b).
\end{enumerate}

\begin{figure}[!t]
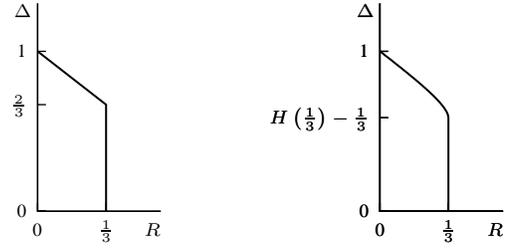

\vspace*{-1em}
\begin{center}
\subfigure[$(p,q,r) = (1/3,1/3,1/3)$]{
\input{memory2.pstex_t}\quad
}\quad
\subfigure[$(p,q,r) = (1/2,1/6,1/3)$]{
\input{memory1.pstex_t}\quad
}
\caption{The optimal $(R,\Delta)$ tradeoff for memory with defective cells.}
\label{fig:graph}
\end{center}
\end{figure}

\end{example}

\medskip

\begin{example} 
\label{ex:binary}
Consider the binary multiplying channel $Y = X \cdot S$, where the
output $Y$ is the product of the input $X \in \{0,1\}$ and the state
$S \in \{0,1\}$.  We assume that the state sequence $S^n$ is drawn
i.i.d.\@ according to $\text{Bern}(\gamma)$.  
It can be easily shown that the optimal tradeoff region
is given by
\begin{align}
R &\le \gamma  H(\alpha) \label{eq:bmc1}\\
\Delta &\le H(\gamma) \label{eq:bmc2}\\
R + \Delta &\le H(\alpha \gamma) \label{eq:bmc3}.
\end{align}
This is achieved by $p(x) \sim \text{Bern}(\alpha)$, independent of
$S$.

As in Example~\ref{ex:memory}(b), there is a tension between the pure
information transmission and the state amplification.  When the goal
is to maximize the pure information rate, we should choose $\alpha^* =
1/2$ to achieve the capacity $C = \gamma$.  But when the goal is to
maximize the state uncertainty reduction rate, we should choose
$\alpha^* = 1$ ($X \equiv 1$) to achieve $\Delta^* = H(\gamma)$.  In
words, to maximize the state uncertainty reduction rate, the
transmitter simply clears the receiver's view of the state.
\end{example}

\section{Extension to Continuous State Space}
\label{sec:gaussian}

\begin{figure*}[th]
\begin{center}
\input{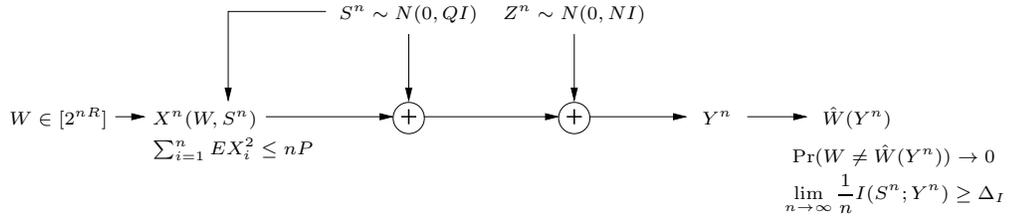}
\caption{Writing on dirty paper.}
\label{fig:wdp}
\end{center}
\end{figure*}

The previous section characterized the tradeoff region $\mathcal{R}^*$
between the pure information rate $R$ and the state uncertainty
reduction rate $\Delta = H(S) - \frac{1}{n} \log |L_n(Y^n)|$.
Apparently the notion of uncertainty reduction rate $\Delta$ is
meaningful only when the channel state $S$ has finite cardinality
(i.e., $|\mathcal{S}| < \infty$), or at least when $H(S) < \infty$.

However, from the proof of Theorem~\ref{thm:noncausal} (the
generalized Fano's inequality in Lemma~\ref{lemma:gen-fano}), along
with the fact that the optimal region is single-letterizable, we can
take an alternative look at the notion of state uncertainty reduction
as reducing the list size from $2^{nH(S)}$ to $|L_n(Y^n)|$.  We will
show shortly in Proposition~\ref{prop:noncausal} that the difference
$\Delta = H(S) - \frac{1}{n} \log |L_n|$ of the normalized list size
is essentially equivalent to the normalized mutual information
$\Delta_I = \frac{1}{n} I(S^n; Y^n)$, which is well-defined for an
arbitrary state space $\mathcal{S}$ and captures the amount of
information the receiver $Y^n$ can learn about the state $S^n$ (or
lack thereof~\cite{Merhav--Shamai2007}).  Hence, the physically
motivated notion $\Delta$ of list size reduction is consistent with
the mathematical information measure $\Delta_I$, and both notions of
state uncertainty reduction can be used interchangeably, especially
when $\mathcal{S}$ is finite.

To be more precise, we define a $(2^{nR}, n)$ code by an encoding
function
\[
X^n: [2^{nR}] \times \mathcal{S}^n \to \mathcal{X}^n
\]
and a decoding function
\[
\hat{W}: \mathcal{Y}^n \to [2^{nR}].
\]
Then the associated state uncertainty reduction rate for the $(2^{nR},
n)$ code is defined as
\[
\Delta_I = \frac{1}{n} I(S^n; Y^n)
\]
where the mutual information is with respect to the joint distribution
\[
p(x^n, s^n, y^n) = p(x^n | s^n) \prod_{i=1}^n p(s_i)
p(y_i|x_i,s_i)
\]
induced by $X^n(W, S^n)$ with message $W$
distributed uniformly over $[2^{nR}]$, independent of
$S^n$.  Similarly, the probability of error is defined as
\[
P_e^{(n)} = \Pr (W \ne \hat{W}(Y^n)).
\]
A pair $(R,\Delta)$ is said to be achievable if there exists a
sequence of $(2^{nR}, n)$ codes with $P_e^{(n)} \to 0$ and
\[
\lim_{n \to \infty} \frac{1}{n} I(S^n; Y^n) \ge \Delta.
\]
The closure of all achievable $(R,\Delta)$ pairs is called the
tradeoff region ${\mathcal{R}}^*_I$.  (Here we use the notation
${\mathcal{R}}^*_I$ instead of $\mathcal{R}^*$ to temporarily
distinguish this from the original problem formulated in terms of the
list size reduction.)

We now show that the optimal tradeoff $\mathcal{R}_I^*$ between the
information transmission rate $R$ and the mutual information rate
$\Delta$ has the same solution as the optimal tradeoff $\mathcal{R}^*$
between $R$ and the list size reduction rate $\Delta$.

\medskip
\begin{proposition}
\label{prop:noncausal}
The tradeoff region ${\mathcal{R}}^*_I$ for a state-dependent channel
$(\mathcal{X} \times \mathcal{S}, p(y|x,s), \mathcal{Y})$ with 
state information $S^n$ noncausally known at the transmitter is the
closure of all $(R, \Delta)$ pairs satisfying
\begin{align}
R &\le I(U; Y) - I(U;S) \tag{\ref{eq:noncausal1}}\\
\Delta &\le H(S) \tag{\ref{eq:noncausal2}}\\
R + \Delta &\le I(X, S; Y) \tag{\ref{eq:noncausal3}}
\end{align}
for some joint distribution of the form $p(s)p(u,x|s)p(y|x,s)$ with
auxiliary random variable $U$. Hence, $\mathcal{R}_I^*$ has the
identical characterization as $\mathcal{R}^*$ in
Theorem~\ref{thm:noncausal}.
\end{proposition}

\begin{IEEEproof}
Let $\mathcal{R}^{**}$ be the region described by
\eqref{eq:noncausal1}--\eqref{eq:noncausal3}. We provide a sandwich
proof $\mathcal{R}^{**} = \mathcal{R}^* \subseteq \mathcal{R}_I^*
\subseteq \mathcal{R}^{**}$, which is given implicitly in the proof of
Theorem~\ref{thm:noncausal}.

More specifically, consider a finite partition%
\footnote{ Recall that the mutual information between arbitrary random
variables $X$ and $Y$ is defined as $I(X; Y) = \sup_{P,Q} I([X]_P;
[Y]_Q),$ where the supremum is over all finite partitions $P$ and $Q$;
see Kolmogorov~\cite{Kolmogorov1956} and Pinsker~\cite{Pinsker1964}.}
to quantize the state random variable $S$ into $[S]$.  Under this
partition, let $\mathcal{R}_{[S]}^{**}$ be the set of all $(R,\Delta)$
pairs satisfying
\begin{align*}
R &\le I(U; Y) - I(U;[S]) \\
\Delta &\le H([S]) \\
R + \Delta &\le I(X, [S]; Y)
\end{align*}
for some joint distribution of the form $p([s])p(u,x|[s])p(y|x,[s])$
with auxiliary random variable $U$. Consider the original list size
reduction problem with state information $[S]$ and let
$\mathcal{R}_{[S]}^*$ denote the tradeoff region. Then
Theorem~\ref{thm:noncausal} shows that $\mathcal{R}_{[S]}^{**} =
\mathcal{R}_{[S]}^*$. In particular, for any $\eps > 0$ and
$(R,\Delta) \in \mathcal{R}_{[S]}^{**}$, there exists a sequence of
$(2^{n(R-\eps)}, 2^{n(\Delta - \eps)}, n)$ codes $X^n(W),
\hat{W}(Y^n), L_n(Y^n)$ such that $P_{e,w}^{(n)} = \Pr(W \ne \hat{W})
\to 0$ and $P_{e,s}^{(n)} = \Pr([S]^n \ne L_n(Y^n)) \to 0$.

Now from the generalized Fano's inequality
(Lemma~\ref{lemma:gen-fano}), the achievable list size reduction
rate $\Delta - \eps$ should satisfy
\[
n(\Delta -\eps) \le I([S]^n; Y^n) + n \eps_n \le I(S^n; Y^n) + n\eps_n
\]
with $\eps_n \to 0$ as $n \to \infty$. Hence by letting $n \to \infty$
and $\eps \to 0,$ we have from the definition of $\mathcal{R}_I^*$
that
\[
\mathcal{R}_{[S]}^{**} = \mathcal{R}_{[S]}^* \subseteq
\mathcal{R}_I^*.
\]
Also it follows trivially from repeating the intermediate steps in the
converse proof of Theorem~\ref{thm:noncausal} that $\mathcal{R}_I^*
\subseteq \mathcal{R}^{**}$.

Finally taking a sequence of partitions with mesh $\to 0$ and hence
letting $\mathcal{R}_{[S]}^{**} \to \mathcal{R}^{**}$, we have the
desired result.
\end{IEEEproof}

\medskip

Since both notions of state uncertainty reduction, the list size
reduction $nH(S) - \log|L_n|$ and the mutual information $I(S^n;Y^n)$,
lead to the same answer, we will subsequently use them interchangeably
and denote the tradeoff region by the same symbol $\mathcal{R}^*$.

\medskip
\begin{example}
\label{ex:wdp}
Consider Costa's writing on dirty paper model depicted in
Figure~\ref{fig:wdp} as the canonical example of a continuous
state-dependent channel.  Here the channel output is given by $Y^n =
X^n + S^n + Z^n$, where $X^n(W,S^n)$ is the channel input subject to a
power constraint $\sum_{i=1}^n E X_i^2 \le nP$, $S^n \sim N(0, QI)$ is
the additive white Gaussian state, and $Z^n \sim N(0, NI)$ is the
white Gaussian noise.  We assume that $S^n$ and $Z^n$ are independent.
\end{example}

For the writing on dirty paper model, we have the following tradeoff
between the pure information transmission and the state uncertainty
reduction.

\medskip
\begin{proposition}
\label{prop:gaussian}
The tradeoff region ${\mathcal{R}}^*$ for the Gaussian channel
depicted in Figure~\ref{fig:wdp} is characterized by the boundary
points $(R(\gamma), \Delta(\gamma)), \enspace 0 \le \gamma \le 1,$
where
\begin{align}
\label{eq:r-gamma}
R(\gamma) &= \frac{1}{2} \log \left( 1 + \frac{\gamma P} {N} \right) \\
\Delta(\gamma)
&= \frac{1}{2} 
\log \left( 1 
+ \frac{\left(\sqrt{Q} + \sqrt{(1-\gamma) P}\right)^2}{\gamma P + N} \right).
\label{eq:delta-gamma}
\end{align}
\end{proposition}
\begin{IEEEproof}[Proof sketch]
The achievability follows from Proposition~\ref{prop:noncausal} with
trivial extension to the input power constraint.  In particular, we
use the simple power sharing scheme proposed
in~\cite{Sutivong--Chiang--Cover--Kim2005}, where a fraction $\gamma$
of the input power is used to transmit the pure information using
Costa's writing on dirty paper coding technique, while the remaining
$(1-\gamma)$ fraction of the power is used to amplify the state.  In
other words,
\begin{equation}
\label{eq:wdp1}
X = V + \sqrt{(1 - \gamma) \frac{P}{Q} }\, S
\end{equation}
with $V \sim N(0, \gamma P)$ independent of $S$, and
\[
U = V + \alpha S
\]
with 
\[
\alpha = \frac{\gamma P}{ \gamma P + N}
\sqrt{\frac{(1 - \gamma) P + Q}{Q}}.
\]
Evaluating $R = I(U;Y) - I(U;S)$ and $\Delta = I(S;Y)$ for each
$\gamma$, we recover \eqref{eq:r-gamma} and \eqref{eq:delta-gamma}.

The proof of converse is essentially the same as that of \cite[Theorem
2]{Sutivong--Chiang--Cover--Kim2005}, which we do not repeat here.
\end{IEEEproof}

As an extreme point of the $(R,\Delta)$, we recover Costa's writing
on dirty paper result
\begin{align*}
C 
&= \half \log \,\biggl( 1 + \frac{P}{N} \biggr)
\end{align*}
by taking $\gamma = 1$. On the other hand, if state uncertainty
reduction is the goal, then all of the power should be used for state
amplification. 
The maximum uncertainty reduction rate
\begin{align*}
{\Delta}^* 
&= \half \log \,
\biggl( 1 + \frac{\bigl(\sqrt{P} + \sqrt{Q}\bigr)^2}{N} \biggr)
\end{align*}
is achieved with $X = \sqrt{\frac{P}{Q}} S$ and $\alpha = 0$.

In \cite[Theorem 2]{Sutivong--Chiang--Cover--Kim2005}, the optimal
tradeoff was characterized between the pure information rate $R$ and
the receiver's state estimation error $D = \frac{1}{n} E ||S^n -
\hat{S}^n(Y^n)||^2$.  Although the notion of
state estimation error $D$ in \cite{Sutivong--Chiang--Cover--Kim2005}
and our notion of the uncertainty reduction rate $\Delta$ appear to
be distinct objectives at first sight, the optimal solutions to both
problems are identical, as shown in the proof of
Proposition~\ref{prop:gaussian}.  There is no surprise here.  Because
of the quadratic Gaussian nature of both problems, minimizing the mean
squared error $E (S - \hat{S}(Y))^2$ can be recast into maximizing the
mutual information $I(S; Y)$, and vice versa.  Also the optimal state
uncertainty reduction rate $\Delta^*$ (or equivalently, the minimum
state estimation error $D^*$ is achieved by the
symbol-by-symbol amplification $X_i = \sqrt{(P/Q)}\, S_i$.

Finally, it interesting to compare the optimal coding scheme
\eqref{eq:wdp1} to the optimal coding scheme when the goal is to
minimize (instead of maximizing) the uncertainty
reduction~\cite{Merhav--Shamai2007}, which is essentially based 
on coherent subtraction of $X$ and $S$ with possible randomization.

\section{Optimal $(R,\Delta)$ Tradeoff: Causal Case}
\label{sec:causal}
The previous two sections considered the case in which the transmitter
has complete knowledge of the state sequence $S^n$ prior to the actual
communication.  In this section, we consider another model in which
the transmitter learns the state sequence on the fly, i.e., the
encoding function
\[
X_i: [2^{nR}] \times \mathcal{S}^i \to \mathcal{X},
\qquad i =1,2,\ldots, n,
\]
depends causally on the state sequence.

We state our main theorem.
\begin{theorem}
\label{thm:causal}
The tradeoff region $\mathcal{R}^*$ for a state-dependent channel
$(\mathcal{X} \times \mathcal{S}, p(y|x,s), \mathcal{Y})$ with 
state information $S^n$ causally known at the transmitter is the union
of all $(R, \Delta)$ pairs satisfying
\begin{align}
R &\le I(U; Y) \label{eq:causal1}\\
\Delta &\le H(S) \label{eq:causal2}\\
R + \Delta &\le I(X, S; Y) \label{eq:causal3}
\end{align}
for some joint distribution of the form $p(s)p(u)p(x|u,s)p(y|x,s)$,
where the auxiliary random variable $U$ has cardinality bounded by
$|\mathcal{U}| \le |\mathcal{X}|\cdot{|\mathcal{S}|}$.
\end{theorem}

As in the noncausal case, the region is convex.  Since the auxiliary
random variable $U$ affects the first inequality \eqref{eq:causal1}
only, the cardinality bound $|\mathcal{U}| \le
|\mathcal{X}|\cdot{|\mathcal{S}|}$ follows again from the standard
argument.  (A looser bound can be given by counting the number of
functions $f: \mathcal{S} \to \mathcal{X}$; see
Shannon~\cite{Shannon1958a}.)  Finally, we can take $X$ as a
deterministic function of $(U,S)$ without decreasing the region.

Compared to the noncausal tradeoff region
$\mathcal{R}^*_\text{nc}$ in Theorem~\ref{thm:noncausal}, the
causal tradeoff region $\mathcal{R}^*_\text{c}$ in
Theorem~\ref{thm:causal} is smaller in general.  More precisely,
$\mathcal{R}^*_\text{c}$ is characterized by the same set of
inequalities~\eqref{eq:noncausal1}--\eqref{eq:noncausal3} as in
$\mathcal{R}^*_\text{nc}$, but the set of joint distributions
is restricted to those with auxiliary variable $U$ independent of $S$.
Indeed, from the independence between $U$ and $S$, we can rewrite
\eqref{eq:causal1} as
\[
R \le I(U;Y) = I(U;Y) - I(U;S) \tag{\ref{eq:causal1}$'$}
\]
which is exactly the same as \eqref{eq:noncausal1}.  Thus the
inability to use the future state sequence decreases the tradeoff
region.  However, only the inequality \eqref{eq:causal1}, or
equivalently, the inequality \eqref{eq:noncausal1}, is affected by the
causality, and the sum rate \eqref{eq:causal3} does not change from
\eqref{eq:noncausal3}.

Since the proof of Theorem~\ref{thm:causal} is essentially identical
to that of Theorem~\ref{thm:noncausal}, we skip most of the steps.
The least straightforward part is the following lemma.

\medskip
\begin{lemma}
\label{lemma:equiv2}
Let $\mathcal{R}$ be the union of all $(R, \Delta)$ pairs satisfying
\eqref{eq:causal1}--\eqref{eq:causal3}.  Let ${\mathcal{R}}_0$ be the
{closure} of the union of all $(R,\Delta)$ pairs satisfying
\eqref{eq:causal1}, \eqref{eq:causal2}, and
\begin{equation}
\label{eq:causal4}
R + \Delta \le I(U, S; Y)
\end{equation}
for some joint distribution $p(s) p(u) p(x|u,s) p(y|x,s)$ where the
auxiliary random variable $U$ has finite cardinality.  Then
\[
\mathcal{R} = {\mathcal{R}}_0.
\]
\end{lemma}
\begin{IEEEproof}[Proof sketch]
The proof is a verbatim copy of the proof of Lemma~\ref{lemma:equiv},
except that here $U$ is independent of $S$, i.e., $p(x,u|s) = p(u)
p(x|u,s)$.  The final step \eqref{eq:inclusion5} follows since
the set of conditional distributions on $X, U = (V,\tilde{U})$
given $S$ of the form
\[
p(x,u|s) = p(v) p(\tilde{u})p(x|v,\tilde{u},s)\tag{$12'$}
\]
with deterministic $p(x|v,\tilde{u},s)$ is as rich as any
$p(\tilde{u})p(x|\tilde{u},s)$, and
\[
I(V, \tilde{U}; Y) \ge I(\tilde{U}; Y) \tag{$13'$}.
\]
With this replacement, the desired proof follows along the same lines
as the proof of Lemma~\ref{lemma:equiv}.
\end{IEEEproof}

As one extreme point of the tradeoff region $\mathcal{R}^*$, we
recover the Shannon capacity formula~\cite{Shannon1958a} for channels
with causal side information at the transmitter as follows:
\begin{equation}
\label{eq:causal-capacity}
C = \max_{p(u)p(x|u,s)} I(U;Y).
\end{equation}
On the other hand, the maximum uncertainty reduction rate $\Delta^*$
for pure state amplification is identical to that for the noncausal
case given in Corollary~\ref{corollary:noncausal}.
\medskip
\begin{corollary}
\label{corollary:causal}
Under the condition of Theorem~\ref{thm:causal}, the maximum
uncertainty reduction rate $\Delta^*$ is given by
\begin{equation}
\label{eq:max-delta3}
\Delta^* = \min \{ H(S), \enspace \max_{p(x|s)} I(X, S; Y) \}.
\end{equation}
\end{corollary}
Thus the receiver can learn about the state essentially at the maximum
cut-set rate, even under the causality constraint. For example, the
symbol-by-symbol amplification strategy $X = \sqrt{\frac{P}{Q}} S$ is
optimal for the Gaussian channel (Example~\ref{ex:wdp}) for both
causal and noncausal cases.

Finally, we compare the tradeoff regions
$\mathcal{R}^*_\text{c}$ and $\mathcal{R}^*_\text{nc}$
with a communication problem that has a totally different motivation,
yet has a similar capacity expression.  In \cite[Situations 3 and
4]{Willems--van-der-Meulen1985}, Willems and van der Meulen studied
the multiple access channel with cribbing encoders.  In this
communication problem, the multiple access channel
$(\mathcal{X}\times\mathcal{S}, p(y|x,s), \mathcal{Y})$ has two inputs
and one output.  The primary transmitter $S$ and the secondary
transmitter $X$ wish to send independent messages $W_s \in
[2^{n\Delta}]$ and $W_x \in [2^{nR}]$ respectively to the common
receiver $Y$.  The difference from the classical multiple access
channel is that either the secondary transmitter $X$ learns the
primary transmitter's signal $S$ on the fly ($X_i(W_x, S^i)$
\cite[Situation 3]{Willems--van-der-Meulen1985}) or $X$ knows the
entire signal $S^n$ ahead of time ($X_i(W_x, S^n)$ \cite[Situation
4]{Willems--van-der-Meulen1985}).  The capacity region $\mathcal{C}$
for both cases is given by all $(R,\Delta)$ pairs satisfying
\begin{align}
R &\le I(X; Y|S) \label{eq:crib1}\\
\Delta &\le H(S) \label{eq:crib2}\\
R + \Delta &\le I(X, S; Y) \label{eq:crib3}
\end{align}
for some joint distribution $p(x,s)p(y|x,s)$.

This capacity region $\mathcal{C}$ looks almost identical to the
tradeoff regions $\mathcal{R}^*_\text{nc}$ and
$\mathcal{R}^*_\text{c}$ in Theorems~\ref{thm:noncausal} and
\ref{thm:causal}, except for the first inequality \eqref{eq:crib1}.
Moreover, \eqref{eq:crib1} has the same form as the capacity
expression for channels with state information available at
\emph{both} the encoder and decoder, either causally or noncausally.
(The causality has no cost when both the transmitter and the receiver
share the same side information; see, for example, Caire and
Shamai~\cite[Proposition 1]{Caire--Shamai1999}.)

It should be stressed, however, that the problem of cribbing multiple
access channels and our state uncertainty reduction problem have a
fundamentally different nature.  The former deals with encoding and
decoding of the signal $S^n$, while the latter deals with uncertainty
reduction in an uncoded sequence $S^n$ specified by nature.  In a
sense, the cribbing multiple access channel is a detection problem,
while the state uncertainty reduction is an estimation problem.

\section{Concluding Remarks}
\label{sec:conc}
Because the channel is state dependent, the receiver is able to learn
something about the channel state from directly observing the channel
output. Thus, to help the receiver narrow down the uncertainty about
the channel state at the highest rate possible, the sender must
jointly optimize between facilitating state estimation and
transmitting refinement information, rather than merely using the
channel capacity to send the state description.  In particular, the
transmitter should summarize the state information in such a way that
the summary information results in the maximum uncertainty reduction
when coupled with the receiver's initial estimate of the state.  More
generally, by taking away some resources used to help the receiver
reduce the state uncertainty, the transmitter can send additional pure
information to the receiver and trace the entire $(R,\Delta)$ tradeoff
region.

There are three surprises here.  First, the receiver can learn about
the channel state and the independent message at a maximum cut-set
rate $I(X,S; Y)$ over all joint distributions $p(x,s)$ consistent with
the given state distribution $p(s)$.  Second, to help the receiver
reduce the uncertainty in the initial estimate of the state (namely,
to increase the mutual information from $I(S;Y)$ to $I(X, S; Y)$), the
transmitter can allocate the achievable information rate $I(U;Y) -
I(U;S)$ in two alternative methods---random binning and its dual,
random covering.  Thirdly, as far as the sum rate $R + \Delta$ and the
maximum uncertainty reduction rate $\Delta^*$ are concerned, there is
no cost associated with restricting the encoder to learn the state
sequence on the fly.


\def\cprime{$'$} \def\cprime{$'$} \def\cprime{$'$}

\end{document}